\documentclass[%
 aip,
 amsmath,amssymb,
 reprint,%
]{revtex4-1}

\usepackage{graphicx}
\usepackage{dcolumn}
\usepackage{bm}

\usepackage[utf8]{inputenc}
\usepackage[T1]{fontenc}
\usepackage{mathptmx}
\usepackage{textgreek}

\usepackage{braket}

\usepackage{color}
\definecolor{col3}{rgb}{0.6, 0.2, 0.6}

\definecolor{col5}{rgb}{0.2, 0.6, 0.1}

\definecolor{col1}{rgb}{0.8, 0.1, 0.1}

\usepackage{xargs}
\newcommandx*{\uin}[1][1=]{_{\text{In}#1}}
\newcommandx*{\uyb}[1][1=]{_{\text{Yb}#1}}
\newcommandx*{\uinn}[1][1=]{_{#1\text{In}}}
\newcommandx*{\uybn}[1][1=]{_{#1\text{Yb}}}
\newcommandx*{\ual}[1][1=]{_{\text{Al}#1}}
\newcommandx*{\uga}[1][1=]{_{\text{Ga}#1}}

\newcommand{\utxt}[1]{_{\text{#1}}}

\DeclareMathOperator{\be}{\beta}
\DeclareMathOperator{\vac}{\ket{\text{vac}}}

\DeclareMathOperator{\xybt}{\xi_{\text{Yb},\tilde{\omega}}}
\DeclareMathOperator{\xint}{\xi_{\text{In},\tilde{\omega}}}

\DeclareMathOperator{\gyb}{\zeta\uyb}
\DeclareMathOperator{\gin}{\zeta\uin}

\DeclareMathOperator{\ovybin}{\braket{\gyb\vert \gin}}
\DeclareMathOperator{\ovinyb}{\braket{\gin\vert \gyb}}
\DeclareMathOperator{\ovinin}{\braket{\gin\vert \gin}}
\DeclareMathOperator{\ovybyb}{\braket{\gyb\vert \gyb}}

\DeclareMathOperator{\ryb}{a^\dagger_\omega}
\DeclareMathOperator{\rin}{b^\dagger_{\omega^\prime}}

\DeclareMathOperator{\dyb}{d^\dagger_\omega}
\DeclareMathOperator{\din}{d^\dagger_{\omega^\prime}}

\DeclareMathOperator{\cyb}{c^\dagger_\omega}
\DeclareMathOperator{\cin}{c^\dagger_{\omega^\prime}}

\draft 

\begin{document}

\title{Photon-mediated entanglement scheme between a ZnO semiconductor defect and a trapped Yb ion} 

\author{Jennifer F. Lilieholm}
\email[]{liliej@uw.edu}
\affiliation{Department of Physics, University of Washington, Seattle, Washington 98195, USA}

\author{Vasilis Niaouris}
\affiliation{Department of Physics, University of Washington, Seattle, Washington 98195, USA}
\author{Alexander Kato}
\affiliation{Department of Physics, University of Washington, Seattle, Washington 98195, USA}
\author{Kai-Mei C. Fu}
\affiliation{Department of Physics, University of Washington, Seattle, Washington 98195, USA}
\affiliation{Department of Electrical and Computer Engineering, University of Washington, Seattle, Washington 98195, USA}

\author{Boris B. Blinov}

\affiliation{Department of Physics, University of Washington, Seattle, Washington 98195, USA}


\date{\today}

\begin{abstract}

We propose an optical scheme to generate an entangled state between a trapped ion and a solid state donor qubit through which-path erasure of identical photons emitted from the two systems. The proposed scheme leverages the similar transition frequencies between In donor bound excitons in ZnO and the $^2P\textsubscript{1/2}$ to $^2S\textsubscript{1/2}$ transition in Yb\textsuperscript{+}. The lifetime of the relevant ionic state is longer than that of the ZnO system by a factor of 6, leading to a mismatch in the temporal profiles of emitted photons. A detuned cavity-assisted Raman scheme weakly excites the donor with a shaped laser pulse to generate photons with 0.99 temporal overlap to the Yb\textsuperscript{+} emission and partially shift the emission of the defect toward the Yb\textsuperscript{+} transition. The remaining photon shift is accomplished via the dc Stark effect. We show that an entanglement rate of 21\,kHz and entanglement fidelity of 94\% can be attained using a weak excitation scheme with reasonable parameters.
\end{abstract}


\maketitle 


Hybrid quantum systems offer the opportunity to combine the benefits of different qubit types while avoiding some of their pitfalls. Task-dependent qubit selection allows the usage of long-lived qubits for memory and qubits with rapid gate speeds for operations. For optical systems, a photon bus can be used to remotely link these systems via photon-heralded entanglement. To successfully generate entanglement, the two different qubit systems must emit identical photons, requiring spectro-temporal engineering of at least one qubit's photon wavepacket. While significant progress has been made toward efficient quantum-frequency conversion~\cite{Siverns2019naw,zaske2012vqf,bock2018hfe,rutz2017qfc}, post-emission temporal photon pulse-shaping~\cite{keller2004cgs, fan2019ssi,karpinski2017bmq} techniques for the narrow-band photons from both trapped ions and solid-state defects is an outstanding challenge.

We have identified two disparate, complementary qubit systems in which high-fidelity photon-mediated entanglement should be possible by direct control over the photon emission process. Trapped ions are a well-studied qubit system with high operational fidelities\cite{Ballance2016} and long coherence times~\cite{wang2017sqq}, but relatively slow initialization and gate speeds~\cite{linke2017ect}. Electron spins in semiconductors have rapid initialization and gate speeds~\cite{warburton2013ssi, fu2008ucd,he2019tqg}, but have shorter coherence times. A hybrid system consisting of ions and electrons bound to donor defects would have the ability to use ions for quantum memory and defects for gate operations, producing a system more rapid and reliable than either qubit alone. 

\begin{figure}
  \centering
 \includegraphics[height=0.55\linewidth]{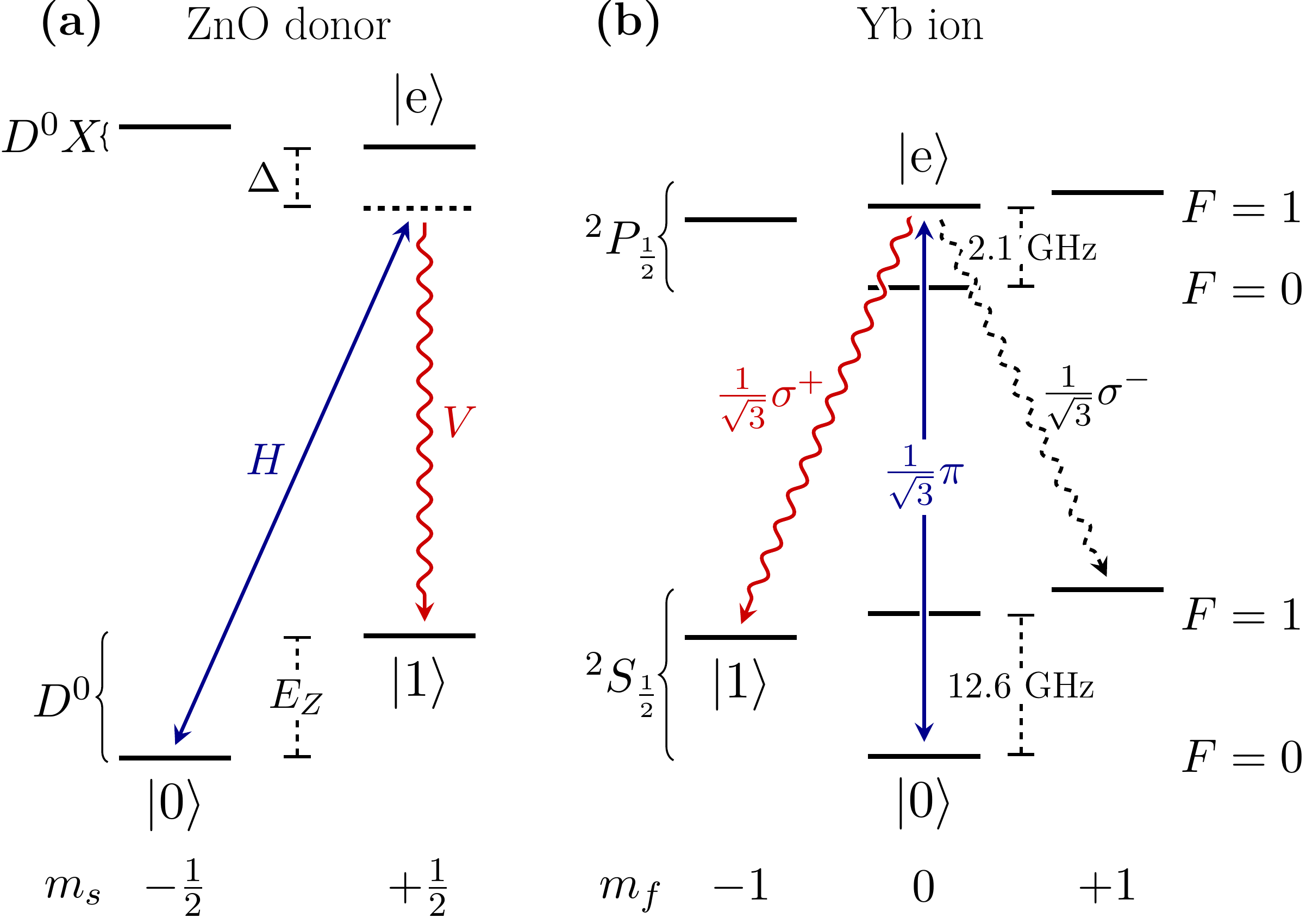}
 \caption{Partial energy level diagrams of ZnO donor (a) and \textsuperscript{171}Yb\textsuperscript{+} (b). Energies are not to scale, except for the $\ket{e}\rightarrow\ket{1}$ transitions. The qubit system in ZnO is comprised of the two electron spins ($\ket{0}$ and $\ket{1}$ with excited level $\ket{e}$) of the neutral donor D\textsuperscript{0}. This state is optically coupled to the donor-bound-exciton state D\textsuperscript{0}X consisting of the donor, two-electron spin singlet, and hole. The \textsuperscript{171}Yb\textsuperscript{+} qubit is formed by the $\ket{F=0, m_{F}=0}$ and $\ket{F=1, m_{F}=-1}$ hyperfine levels in the $^2$S\textsubscript{1/2} ground state. 
 \label{trans}}
 \end{figure}

 Yb\textsuperscript{+} and the ZnO donor were chosen as the target systems for their shared transition near 369\,nm: the $^2P\textsubscript{1/2}$ to $^2S\textsubscript{1/2}$ transition in \textsuperscript{171}Yb\textsuperscript{+} and the In neutral donor bound exciton (D$\textsuperscript{0}$X) to neutral donor (D$\textsuperscript{0}$) transition in ZnO (Fig.\,\ref{trans}). In:ZnO is analogous in structure to the better-known P:Si qubit system~\cite{kane1998sbn}, however ZnO is a direct band gap semiconductor enabling efficient donor coupling to photons.  While the two transition frequencies are quite close ($\delta$ = 0.36~THz), the excited state lifetimes differ by a factor of  6 resulting in a large temporal mismatch.   Prior semiconductor spin - trapped ion entanglement schemes addressed similar temporal mismatch by using coherent scattering~\cite{waks2009phe} or sacrificing fidelity~\cite{meyer2015dpc}. Here we demonstrate that pulse shaping can be a powerful tool to attain high-fidelity entanglement and show that an entanglement rate of 21\,kHz and fidelity of 94\% is feasible.   


A heralded entanglement scheme based on weak excitation, single-photon detection and which-path erasure can be used to entangle the two systems, similar to the proposal by Cabrillo \textit{et al.}~\cite{cabrillo1999ces} Fig.\,\ref{trans} depicts the relevant energy levels and excitation/decay pathways for the donor and ion.  Here D\textsuperscript{0} system is in the Voigt (B$\perp \hat k$) geometry but the Faraday geometry could also be utilized. The donor is coupled to an optical cavity detuned by $\Delta$ from the D\textsuperscript{0}X-D\textsuperscript{0} transition. 

\begin{figure}
 \includegraphics[width=7.5cm]{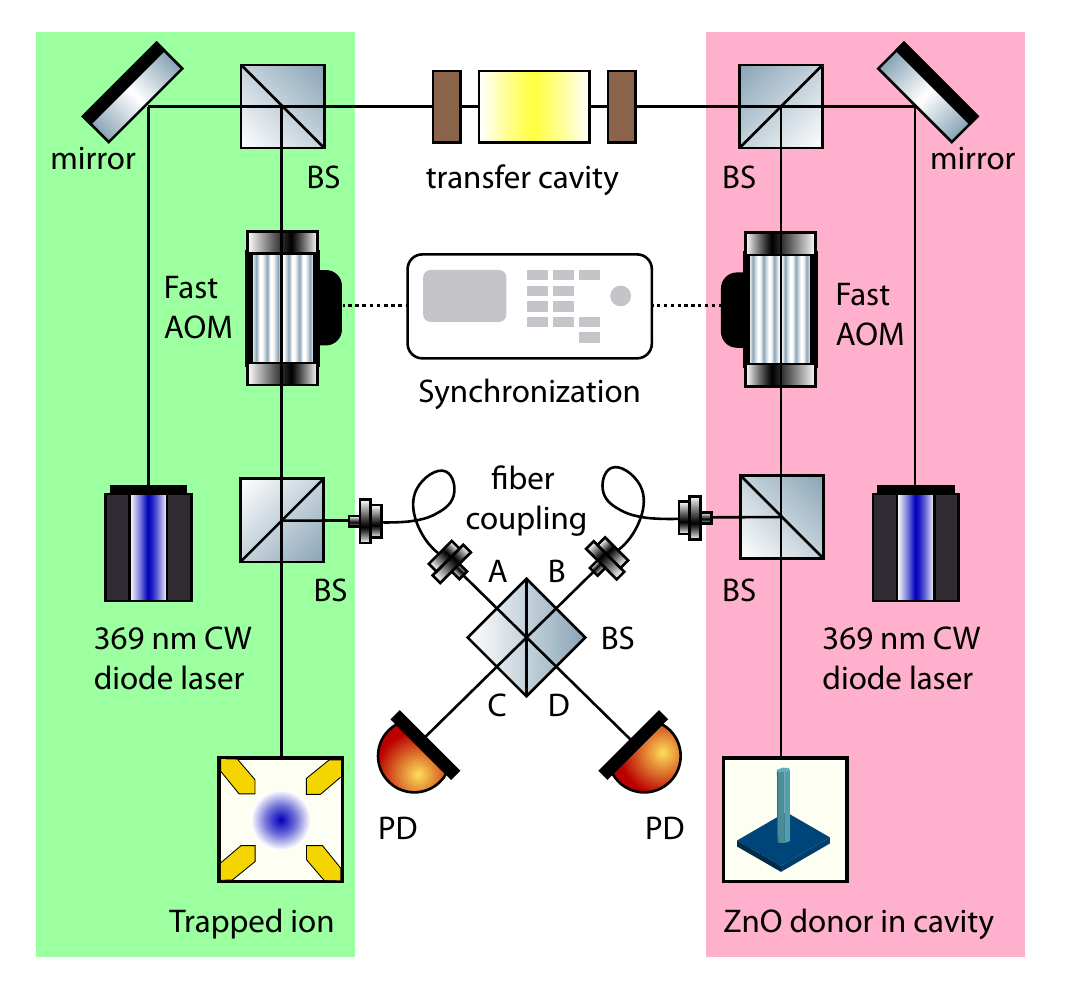}
 \caption{A schematic for remote entanglement of a trapped ion qubit (left) and donor qubit in ZnO (right). A transfer cavity phase-locks the two 369\,nm excitation lasers. The two acousto-optic modulators (AOM) are synchronized and programmed to output the calculated pulse shapes for their respective qubits. Photons collected from the two qubits interfere on the central beam splitter (BS) via inputs A and B. Successful entanglement is heralded by the detection of a single photon by photodetectors (PD) at outputs C, D.  For the trapped ion, a B-field pointing along the direction of fluorescence collection breaks the degeneracy between the $\ket{F=1}$ states and defines the quantization axis. For the donor, the B-field points either parallel to the direction of emitted photons (Faraday geometry) or perpendicular to it (Voigt geometry).\label{ent}}
 \end{figure}

The diagram of the experiment is shown in Fig.\,\ref{ent}. The Yb\textsuperscript{+} and In donor are first initialized using optical pumping to $\ket{F=0,m_F=0}$ and $\ket{m_s=-1/2}$, respectively, producing the initial state $|\Psi\rangle_i = |0\rangle_{\textrm{Yb}}\otimes|0\rangle_{\textrm{In}}\otimes |\textrm{vac}\rangle  \equiv |0;0;\textrm{vac}\rangle$. Next, each system is excited to $\ket{e}$\textsubscript{In} or $\ket{e}$\textsubscript{Yb}, using resonant or near-resonant pulsed excitation. Here, we assume the weak excitation limit (excitation probability $p_{1,x}<10\%$, $x=\{\text{Yb,In}\}$).

The state of the ZnO donor and ion is now given by

\begin{equation}
    \begin{split}
      \ket{\Psi}_{c} 
        = \be_1 \ket{0;0;\text{vac}}
      & + \be_4 \ket{1;1;\gyb,\gin} +  \\
        + \be_2 \ket{0;1;\gin}
      & + \be_3 \ket{1;0;\gyb},
    \end{split}
\end{equation}
where the emitted photons on paths A and B of Fig.~\ref{ent} $\ket{\gyb}=\sum_\omega \xi\uyb[,\omega] \ryb \vac$ and $\ket{\gin}=\sum_{\omega^{\prime}} \xi\uin[,\omega^{\prime}] \rin \vac$ are given by a sum over all modes $\omega$ ($\omega^\prime$) with coefficients $\xi\uyb[,\omega]$ ($\xi\uin[,\omega^\prime]$) and creation operators $\ryb$ ($\rin$). 
The coefficients $\beta$ emerge from the excitation ($p_{1,x}$)  probabilities of the two systems, the phase gained from excitation laser phases ($\phi_{x,L}$), and the distance travelled by the collected photon ($\phi_{x,d}$):

\begin{equation}
    \begin{split}
\beta_1 =\sqrt{(1-p\uybn[1,])(1-p\uinn[1,])}e^{i(\phi\uyb[,L]+\phi\uin[,L])}\\
    \beta_2  =  \sqrt{p\uinn[1,](1-p\uybn[1,])}e^{i(\phi\uyb[,L]+\phi\uin[,d])}\\
    \beta_3=\sqrt{p\uybn[1,](1-p\uinn[1,])}e^{i(\phi\uyb[,d]+\phi\uin[,L])}\\
    \qquad
    \beta_4=\sqrt{p\uinn[1,]p\uybn[1,]}e^{i(\phi\uyb[,d]+\phi\uin[,d])}\\
    \end{split}
\end{equation} 

By phase locking the laser pulses, we can ignore $\phi_{x,L}$. 

Collected photons from both systems interfere on the beamsplitter, which erases which-path information.
 
Entanglement is heralded by the detection of a single photon at one of the two photodetectors. With the appropriate choice for $p\uybn[1,]$, $p\uinn[1,]$, and the collection efficiency from each system (supplemental material), photon detection in path D projects the ion-donor qubits onto the renormalized entangled state
\begin{equation}
    \begin{split}
        \ket{\Psi} = \frac{1}{\sqrt{2}}\left( \ket{0;1;\gin}-i e^{i\Delta\phi} \ket{1;0;\gyb} \right),
    \end{split}
\end{equation}
where $\Delta\phi$ is determined by the optical path length difference. 
Similar expression can be derived for detector C. Tracing over all photon modes, we get the reduced Yb$^+ -$ In density matrix

\begin{equation}
    \begin{split}
        \rho^{\text{Yb,In}} 
         & = \frac{1}{2}  \ket{0;1} \bra{0;1}
         + \frac{1}{2}  \ket{1;0} \bra{1;0} +\\
         & + \frac{1}{2} \left(i e^{i\Delta\phi} \ovybin \ket{0;1} \bra{1;0}
         +c.c.\right),
    \end{split}
\end{equation}
where  $\ovybin = \sum_{\tilde{\omega}} \xybt^* \xint$ is the overlap of the photons from the Yb\textsuperscript{+}  and ZnO systems.

Factors which affect the entanglement fidelity are photon overlap, 
false identification of both-system excitation as a single-system excitation, and atomic recoil from the ion interacting with the excitation laser. Accounting for these sources of error, the final fidelity is:
 \begin{equation}
F=\frac{1}{2+c_{1}^{2}}[1+F\utxt{dyn}\text{Re}(\ovybin)]
\label{eq:fidelity}
\end{equation}
where c\textsubscript{1} depends on the excitation probabilities and detection efficiencies of both systems (supplemental material) and $F\textsubscript{dyn}$ is related to the photon recoil~\cite{cabrillo1999ces}. Motion of the trapped ion due to photon recoil during the absorption/emission process can shift the frequency of the photon and reduce fidelity of the entangled state. Note that for the ZnO donor, absorption/emission are recoilless due to the Mössbauer effect.
For a Doppler-cooled $^{171}$Yb$^{+}$ in a 1~MHz trap in geometry where the ion is excited by a laser pulse parallel to the light collection direction, the expected  $F\utxt{dyn}$ is 96\%~\cite{cabrillo1999ces}. In addition, uncertainty in both $\phi_{x,L}$ and $\phi_{x,d}$ leading to an undesired phase factor $e^{i\epsilon}$ between the terms in Eq. ~3 can further degrade the fidelity according to $\text{Re}(\ovybin) \rightarrow \text{Re}(e^{i\epsilon} \ovybin).$
Other factors that may further decrease the fidelity include photodetector dark counts, background luminescence from ZnO, and D$\textsuperscript{0}$X spectral diffusion~\cite{Humphreys2018,slodivcka2013aae}.


Photon collection efficiency primarily affects the protocol’s probability of success. For trapped ions, light collection is challenging due to the high-vacuum environment and the need to isolate ions from decoherence-inducing surfaces. Typical light collection efficiency is 2-4\% utilizing off-the-shelf long working distance microscope objectives~\cite{blinov04nat}, while optics based on in-vacuum lenses~\cite{araneda2018prl} and custom high-NA objectives~\cite{stephenson2020hrh} are capable of collecting up to 10\% of the emitted photons. Further enhancement is possible by integrating a metallic parabolic mirror as an RF electrode of the ion trap~\cite{chou2017nsi}. Ions are trapped at the focus of the mirror, so that the emitted photons are collimated upon reflection from the mirror with an expected 32\% overall coupling efficiency into a single-mode optical fiber.  As we show below, the parabolic mirror trap also provides a mechanism for polarization filtering. Longer term, integrated-photonics platforms may provide a path toward high-NA collection from scalable arrays of ions~\cite{bruzewicz2019tiq}.
 
\begin{figure}
 \includegraphics[width=0.9\linewidth]{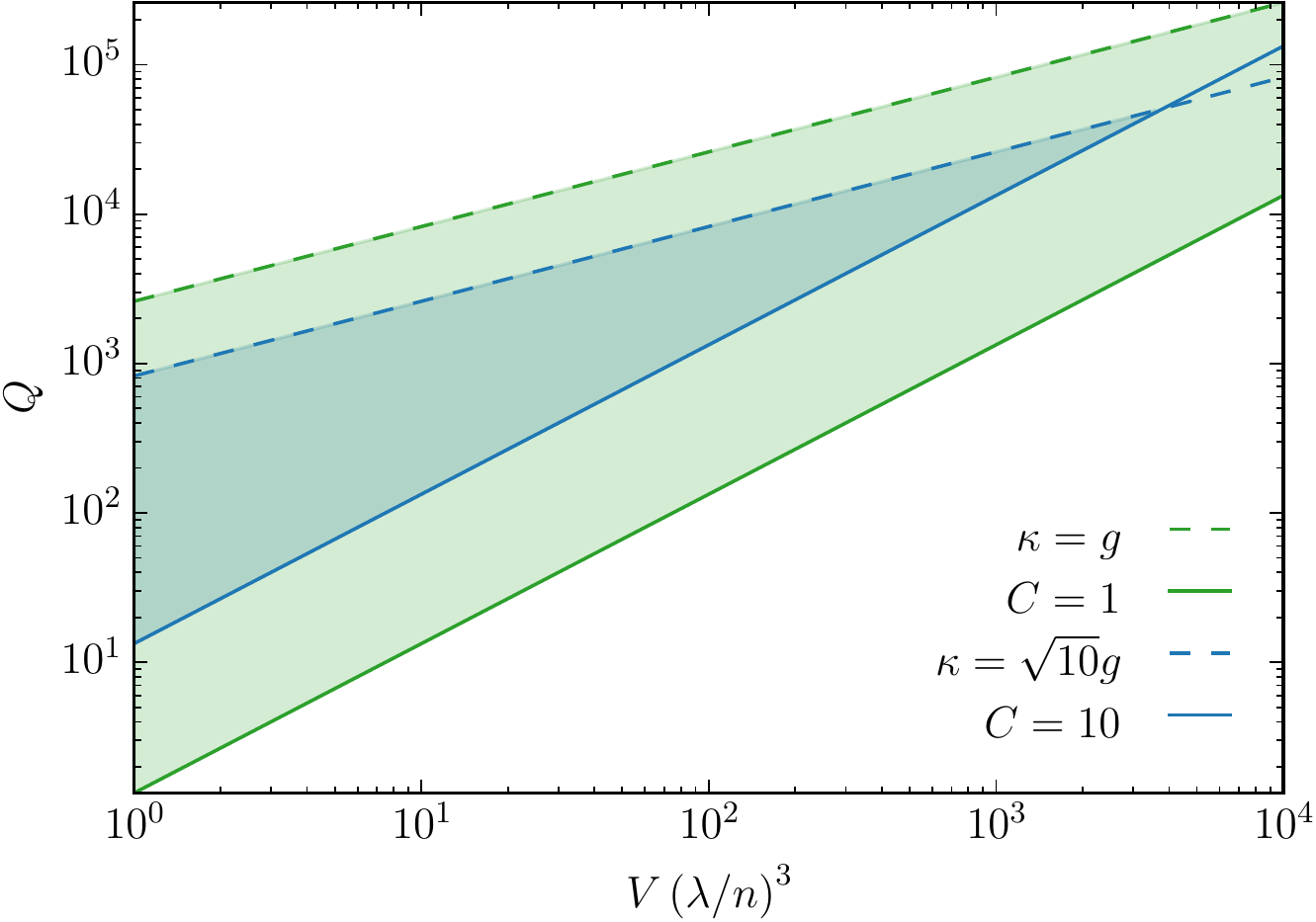}
 \caption{ZnO cavity parameter space ($\kappa$, $g$, $C$) satisfying the photon pulse-shaping requirements in terms of the quality factor $Q$ and the mode volume $V$. The green area corresponds to $C\geq1$ and $g\leq\kappa$, and the blue area corresponds to $C\geq10$ and $\sqrt{10}g\leq\kappa$. \label{QV}}
 \end{figure}

For the donor, a photonic cavity can be fabricated in ZnO to enhance collection efficiency. As shown in Fig.\,\ref{QV}, cavities which satisfy high cooperativity $C={g^2}/{\kappa\Gamma\uin}$ (here $g$ is the donor-cavity coupling strength, $\kappa$ is the cavity decay rate and $\Gamma\uin$ is the spontaneous decay rate) in the “bad cavity” limit necessary for the pulse-shaping procedure described below, lie in a band of readily achievable $Q/V$ ratios with today’s nanophotonic fabrication techniques (here $Q$ is the quality factor and $V$ is the mode volume of the cavity). Due to intrinsic band-edge absorption, the high quality factor region in Fig.\,\ref{QV} may not be achievable at D\textsuperscript{0}X-D\textsuperscript{0} transition~\cite{nur2019spc}, thus low mode volume cavities with moderate quality factors should be targeted. While nanophotonic fabrication in ZnO is relatively immature compared to other quantum defect host crystals, small mode volume ZnO nanowire cavities have enabled UV lasers~\cite{huang2001rtu} and ZnO cavities fabricated by focused ion beam milling~\cite{chang2016zob}, a method that has been used to achieve high cooperativity in rare-earth doped systems~\cite{zhong2018oas}, exhibit quality factors up to 1000. In the limit that the cavity photon loss rate $\kappa$ is dominated by coupling to the output mode, over 50\% collection efficiency into a waveguide for planar geometry cavities~\cite{arcari2014nuc} or into an objective lens for nanowire cavities~\cite{senellart2017hps} is possible. 

As shown in Eq.\,\ref{eq:fidelity}, for high fidelity entanglement, the frequency, polarization, and temporal shape of the photons emitted by the two systems must be matched to maximize $\text{Re}(\ovybin)$. The type of donor used affects the amount of frequency shift required to match the emission frequency of Yb\textsuperscript{+}. Of the three primary donor candidates, Al, Ga and In, the In D\textsuperscript{0}X transition is closest to the Yb$^+$ transition, 
$v\uin=v\uyb + 0.36$~THz~\cite{meyer2004bed}, where $v\uin$ and $v\uyb$ are the values of the $\ket{0}\rightarrow \ket{e}$ transitions with zero magnetic field, and in the absence a DC Stark shift. 
The donor will be integrated in an optical cavity detuned from the relevant transition by $\sim$200\,GHz. The remaining frequency shift will be attained via the DC Stark effect.  Electric field tuning in a similar quantum dot trion system has shown that several meV of tuning is possible~\cite{bennett2010gse}.

Decay from $\ket{\text{e}}$\textsubscript{Yb} ($^2P_{1/2}$ $\ket{F=1, m_F=0}$) can occur along three different channels, producing either a $\sigma^{\pm}$ Raman photon or a $\pi$ Rayleigh photon (see Fig.\ref{trans}). A pure polarization state is required for polarization matching with the photon emitted by the ZnO donor. While the use of a high-NA collection optic increases the photon collection efficiency, it can pose problems for polarization purity. However, the parabolic mirror can be utilized to filter out the undesired $\pi$ polarized photons when the optical axis is oriented along the quantization axis defined by the applied magnetic field~\cite{kim2011ecs}. In this geometry, the $\pi$-polarized photons reflected off the mirror have a radial polarization pattern, which completely destructively interferes when focused into a single-mode optical fiber. The $\sigma$-polarized photons, on the other hand, have an elliptical polarization upon reflection from the mirror. The eccentricity increases with radial distance from the center, with perfectly circular polarization at the center of the reflected beam and linear polarization at the edge. The linear component is filtered out by destructive interference in the optical fiber. 

  \begin{figure}[tb!]
 \centering
 \includegraphics[height=0.4865\linewidth]{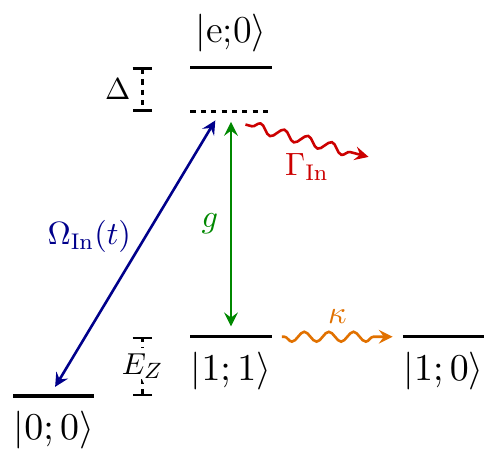}
 \caption{ZnO energy-level system used in pulse-shaping calculations. The kets represent the In:ZnO state and the associated photon number. 
\label{pl}}
 \end{figure}
  \begin{figure*}
\includegraphics[width=0.9\textwidth]{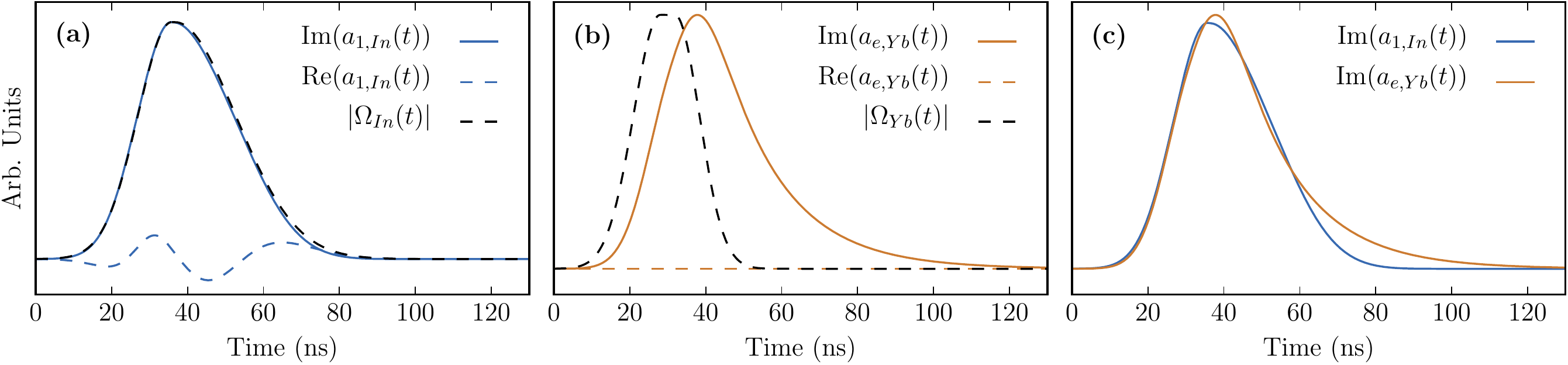}
  \caption{(a) Excitation pulse and temporal wavefunction of the emitted photon for the ZnO system. The parameters used are $\Delta=2\pi\times$(200\,GHz), $\sigma_{1}=8.9$\,ns, $\sigma_{2}=16$\,ns, $\tau=35.8$\,ns, $t_{h}=0.85$\,ns, $\Omega_{max}=2\pi\times$(2.9\,GHz), $\theta_1=2\pi\times$(6.9\, MHz), $\theta_0=2\pi\times(-0.15)$, $g=2\pi\times(15$\,GHz), and $\kappa=2\pi\times$(60\,GHz). (b) Excitation pulse and temporal wavefunction of emitted photon for the Yb\textsuperscript{+} system with $\sigma_1=7.0$\,ns, $\sigma_2=6.4$\,ns, $\tau=28$\,ns, $t_{h}=3.9$\,ns, $\Omega_{max}=2\pi\times$(8.1\,MHz), $\theta_1=0$\, GHz, and $\theta_0=2\pi\times(0.50)$. (c) Imaginary parts of both wavefunctions, leading to $\text{Re}(\ovybin)\simeq0.99$.} \label{pulses}
 \end{figure*}
In the Voigt geometry, with the applied magnetic field perpendicular to the crystal axis, the branching ratio between the ZnO donor Raman transitions $\ket{\text{e}}\uin \rightarrow \ket{0}\uin$ 
and $\ket{\text{e}}\uin \rightarrow \ket{1}\uin$ is approximately 1:1~\cite{linpeng2018cps,wagner2009gvb}. 
For a cavity with large $V$ and high $Q$ (e.g. ring resonator~\cite{liu2018uhq}), the cavity resonance will be narrower than the Zeeman splitting of D$^0$, allowing for selective coupling of the desired Raman transition. For high $V$, the size of the cavity is large compared to the excitation beam diameter, so polarization selection can be attained by 
selectively exciting a small area of the cavity, where only one dipole moment is coupled to the cavity mode.
For cavities with low $Q$ and $V$, polarization and frequency selection can be achieved via cross  polarization~\cite{meyer2015dpc}, waveguide excitation~\cite{huber2020ffs} and spectral filtering.

Matching the temporal profiles of the emitted photons poses a greater challenge. The 
$^2$P\textsubscript{1/2} Yb\textsuperscript{+} state lifetime is 8.1\,ns~\cite{olmschenk2009ml6}, while that of D\textsuperscript{0}X state in ZnO is only 1.4\,ns~\cite{wagner2011bez}. Post-emission pulse shaping~\cite{wright2017ssq,baek2008tsh} is not feasible because the ZnO and Yb photons are too narrow band for these dispersive methods. Instead, the photons emitted by the ZnO donor can be pulse-shaped at their creation~\cite{law1997dgb} by modulating the intensity of the excitation pulse. The ZnO cavity is constructed with parameters within the ``bad cavity" regime ($\kappa \gg g^2/ \kappa \gg \Gamma\uin$)~\cite{law1997dgb}. The large cavity decay rate ensures that we are not in the strong coupling regime, so the donor excitation follows the optical pulse, while the high cooperativity ensures that the donor decays via Raman emission into the cavity.


While it is possible to obtain an analytic expression for the ideal excitation pulse shape for maximum photon overlap~\cite{vasilev2010spm}, in this work we limit ourselves to experimentally attractive Gaussian pulses and performed numerical simulations to determine photon temporal overlap,  given the practical cavity considerations discussed above. The donor defect is modeled as a three level system with initial state $\ket{0}\uin$ (Fig. \ref{pl}) connected to the excited state $\ket{\text{e}}\uin$ by an excitation pulse of Rabi frequency $\Omega\uin(t)$ and detuning $\Delta$. We neglect the effect of the other excited state level. 
The cavity is coupled to the $\ket{e;0}\leftrightarrow\ket{1;1}$ transition with detuning $\Delta$ and coupling strength $g$. Photons from this transition have a spontaneous radiative decay rate of $\Gamma_{\text{In}}$. 
Photons escape the cavity at the cavity decay rate $\kappa$. 
The equations of motion for the population amplitudes are~\cite{law1997dgb,vasilev2010spm}

\begin{equation} \label{3leveleqz}
i \frac{d}{dt}a\uin(t)=
\frac{1}{2}\begin{pmatrix}
0 & \Omega\textsubscript{In}(t) & 0\\
\Omega\textsuperscript{*}\textsubscript{In}(t) & 2\Delta - i \Gamma_{\text{In}} & 2g \\
0 & 2g & -i\kappa
\end{pmatrix}
a\uin(t),
\end{equation}
where $a\uin(t)=[a\uinn[0,](t), a\uinn[\text{e},](t), a\uinn[1,](t)]^T$.

The Yb\textsuperscript{+} is modeled in a similar manner but without a cavity. The ground state $\ket{0}\uyb$ is coupled to the excited state $\ket{\text{e}}\uyb$ by the Rabi pulse $\Omega\uyb(t)$. Decay from the excited state occurs with the rate $\Gamma_{\text{Yb}}$. The equations of motion are:
\begin{equation} \label{3leveleqy}
i \frac{d}{dt} 
\begin{pmatrix}
a\textsubscript{0,Yb}(t) \\
a\textsubscript{e,Yb}(t) \\
\end{pmatrix}=
\frac{1}{2}\begin{pmatrix}
0 & \Omega\textsubscript{Yb}(t)\\
\Omega\textsuperscript{*}\textsubscript{Yb}(t) &  - i \Gamma_{\text{Yb}}\\
\end{pmatrix}
\begin{pmatrix}
a\textsubscript{0,Yb}(t) \\
a\textsubscript{e,Yb}(t) \\
\end{pmatrix}
\end{equation}

The emission rates of the photons from the ZnO and Yb\textsuperscript{+} systems are $\kappa|a\uinn[1,](t)|^2$ and $\Gamma\uyb|a\uybn[e,](t)|^2$, respectively~\cite{law1997dgb}, with temporal wavefunctions given by normalizing the population amplitudes $a\textsubscript{1,In} (t)\rightarrow A\textsubscript{1,In}(t)$ and $a\textsubscript{e,Yb}(t)\rightarrow A\textsubscript{e,Yb}(t)$. By controlling the Rabi frequencies $\Omega\uin (t)$ and $\Omega\uyb (t)$, it is possible to engineer the real component of the photon overlap $\int_{-\infty}^{\infty}A^{*}\uybn[e,](t)A\uinn[1,](t)dt=\ovybin$ to \textasciitilde 0.99 for practical experimental parameters using the control pulses shown in Fig.\,\ref{pulses}. The optimized pulse is restricted to a Gaussian pulse shape with adjustable rise time $\sigma_{1}$, fall time $\sigma_{2}$, time to pulse max $\tau$, hold time $t_{h}$, maximum pulse height $\Omega_{max}$, and phase factor e\textsuperscript{$i \alpha(t)$} where $\alpha(t)=\theta_0+\theta_1 t$ describes a linear time-dependent phase. Setting either pulse to achieve a desired excitation probability $p_{1,x}$, we iteratively sweep the pulse parameters for the other system to obtain local maxima in the overlap.  

The probability of successful entanglement is
\begin{equation}
    P_{\text{succ}}=[p\uybn[1,]p\uybn[2,](1-p\uinn[1,])
    +p\uinn[1,]p\uinn[2,](1-p\uybn[1,])]\eta
\end{equation}
where $p_{2,x}$ is the collection efficiency from each system, and $\eta$ is the quantum efficiency of the detector, which can be as high as \textasciitilde 80\% using superconducting nanowire single photon detectors (SNSPD's)~\cite{Crain2019} for photons at 369\,nm. With a parabolic mirror ion trap, collection efficiency for Yb\textsuperscript{+} systems is 32\%; the ZnO system is set to 34\% collection efficiency to match the coefficients in Eq.~2 to create a maximally entangled state. Excitation probabilities depend on the pulse shaping requirements, and need to be kept low (<10\%) to minimize error. For good fidelity while still maintaining a reasonably high success probability, we use excitation probabilities around 5\%.
 
Each experimental run begins with \textasciitilde1\,\textmu s of optical pumping, followed by the \textasciitilde 10\,ns  excitation pulse. If a single photon is detected, then the state readout is performed, taking \textasciitilde10\,\textmu s and limited by the ion~\cite{Crain2019}. We find a success probability of \textasciitilde2.7\%, leading to an entanglement generation rate of 21\,kHz. Practically, this rate will be further decreased by 
the interferometer phase stabilization and defect frequency stabilization steps~\cite{Humphreys2018}.

With all experiments using this type of protocol,  there is a tradeoff between success probability and fidelity~\cite{slodivcka2013aae,Humphreys2018}. One can always increase the success probability by increasing the excitation probability, but this degrades the fidelity according to Eq \ref{eq:fidelity}. Further, in order to be useful, the entanglement rate needs to be comparable to the rate of decoherence. While the demonstrated coherence time for trapped ytterbium ions is long~\cite{Wang2017} (10 minutes), the spin echo time T\textsubscript{2} of ensemble donor bound excitons in ZnO is only 50\,\textmu s. However, the fundamental limit of T\textsubscript{2} is the longitudinal spin relaxation time T\textsubscript{1} which exceeds 100~ms~\cite{linpeng2018cps} and may allow for improvement through chemical and isotope purification~\cite{tribollet2009ten}.



In summary, a ZnO donor defect qubit and a single trapped Yb\textsuperscript{+} ion can be remotely entangled via a photonic link at 369\,nm. Pulse shaping techniques can be used to alter the temporal profile of the photon emitted by the donor to attain the temporal wavefunction overlap of 0.99 with the photon emitted by the trapped ion, leading to an entangled state fidelity of 94\% with realistic parameters.  

\section*{Supplementary Material}
See supplemental material for a derivation of the fidelity expression (Eq. \ref{eq:fidelity}).

\section*{Acknowledgement}

We thank Xiayu Linpeng for assistance with creating Fig.\,\ref{ent}. This material is based upon work supported by the U.S. Department of Energy, Office of Science, Office of Basic Energy Sciences under award DE-SC0020378.

\section*{AIP Publishing Data Sharing Policy}

The data that supports the findings of this study are available within the article.

\bibliography{References}


\section{Supplemental Material}
\subsection{Maximally entangled state and fidelity}

We first define the state$\ket{\Psi_{1}}$ of $\text{Yb}^{+}$ and the state $\ket{\Psi_{2}}$ of the In donor. We begin by optically pumping both systems into the ground state
\begin{subequations} 
\begin{align}
    \ket{\Psi_{1}}& = \ket{0}\uyb \tag{S1a}\\
    \ket{\Psi_{2}}& = \ket{0}\uin. \tag{S1b}
\end{align}
\end{subequations}
We now apply an excitation pulse to both species with $p_{1,x}\ll 1$ so that the probability of both systems being excited during the same experimental run is small.  

The states of both systems are given by:
\begin{equation}
\ket{\Psi_{1}}=\sqrt{p\uybn[1,]}e^{i\phi_{D_1}}\ket{1}\ket{\zeta\uyb}+\sqrt{1-p\uybn[1,]}e^{i\phi_{L_1}}\ket{0}\vac
\tag{S2}
\end{equation}
\begin{equation}
\ket{\Psi_{2}}=\sqrt{p\uinn[1,]}e^{i\phi_{D_2}}\ket{1}\ket{\zeta\uin}+\sqrt{1-p\uinn[1,]}e^{i\phi_{L_2}}\ket{0}\vac
\tag{S3}
\end{equation}
where $\phi_{L_x}$ denotes the phase of the laser at species x, $\phi_{D_x}$ denotes the phase of the emitted photon after travelling a distance $D_{x}$, $\vac$ is the vacuum state, and $\ket{\zeta\uyb}=\sum_\omega \xi\uyb[,\omega] \ryb \vac$ and $\ket{\zeta\uin}=\sum_{\omega^{\prime}} \xi\uin[,\omega^{\prime}] \rin \vac$ are the temporal wavefunctions of emitted photons from each system. The temporal wavefunctions are given by a sum over all modes $\omega$ ($\omega^\prime$) with coefficients $\xi\uyb[,\omega]$ ($\xi\uin[,\omega^\prime]$) and raising operators $\ryb$ ($\rin$). We phase lock the laser systems to set $\phi_{L_1}=\phi_{L_2}=0$.
Assuming we collect a single photon with efficiency $p_{2,x}$ from either system, we obtain the state  (not normalized)
\begin{equation}
\begin{split}
\ket{\Psi_{1,2}}=
\sqrt{p\uybn[1,](1-p\uinn[1,])p\uybn[2,]}\ket{1,0}\sum_\omega \xi\uyb[,\omega] \ryb \vac \\
+\sqrt{p\uinn[1,](1-p\uybn[1,])p\uinn[2,]}e^{i\Delta\phi}\ket{0,1}\sum_{\omega^{'}}^{} \xi\uin[,\omega^{'}] \rin \vac
\end{split}
\tag{S4}
\end{equation}
 where $\Delta\phi=\phi_{D_2}-\phi_{D_1}$ is the difference in optical path length between the two qubit systems and we have dropped the terms $\ket{1,1}$ and $\ket{0,0}$, since they will eventually be projected out upon the detection of a single photon.

Here, we note that to obtain a maximally entangled state we want to set
\begin{equation}
p\uybn[1,](1-p\uinn[1,])p\uybn[2,]=p\uinn[1,](1-p\uybn[1,])p\uinn[2,].
\tag{S5}
\end{equation}

Since we use $p_{1,x}$ to achieve good temporal overlap, and typically $p\uybn[2,]<p\uinn[2,]$, this is accomplished by lowering $p\uinn[2,]$.

Now, at the beamsplitter we choose the transformation $\ryb \rightarrow (\cyb + i \dyb)/\sqrt{2},\;
    \rin \rightarrow (\din + i \cin)/\sqrt{2}$
,
where $\ryb$, $\rin$ are the raising operators of the respective paths A and B and $\cyb$, $\dyb$ are the raising operators in paths C and D, as depicted in Fig, 2 of main text.

We also have that $i(\phi_{D_2}-\phi_{D_1})=i\Delta\phi$ where $\Delta\phi$ is the difference in phase between photons traversed from each system. To account for the reflection of one of the two paths in the beamsplitter, we set a phase difference of $\frac{\pi}{2}$ between $\ket{1,0}$ and $\ket{0,1}$ states.
 
We then obtain the entangled state upon detection of a single photon 
\begin{equation}
\begin{split}
\ket{\Psi_{1,2}}&=\frac{1}{\sqrt{2}} [
\ket{1,0}\sum_\omega \xi\uyb[,\omega] \frac{\cyb+i\dyb}{\sqrt{2}} \vac  \\
&  -ie^{i\Delta\phi}\ket{0,1}\sum_{\omega^{'}}^{} \xi\uin[,\omega^{'}] \frac{\din+i\cin}{\sqrt{2}} \vac ]
\end{split}
\tag{S6}
\end{equation}

The density matrix can then be computed. Let us first assume the photon was detected on path D, and not on path C. Tracing over photon states in the path D, and over all photon frequencies $\omega$, we obtain
\begin{equation}
\begin{split}
\rho^{\text{Yb,In,D}}=\frac{1}{4}& [ \sum_\omega  \xi^{*}\uyb[,\omega]\xi\uyb[,\omega]
\ket{1,0}\bra{1,0}  \\
+ &\sum_{\omega}\xi^{*}\uyb[,\omega]\xi\uyb[,\omega]
\ket{0,1}\bra{0,1}\\
-ie^{i\Delta\phi}&\sum_\omega \xi^{*}\uyb[,\omega]\xi\uin[,\omega]  \ket{0,1}\bra{1,0} \\
 +ie^{-i\Delta\phi} &  \sum_{\omega}^{} \xi^{*}\uin[,\omega]\xi\uyb[,\omega] \ket{1,0}\bra{0,1} ]
\end{split}
\tag{S7}
\end{equation}

The same matrix can be found for the path C. Summing the density matrices we then find the complete density matrix including paths C and D
\begin{equation}
\begin{split}
\rho^{\text{Yb,In}}=\frac{1}{2}( 
\ket{1,0}\bra{1,0}+
\ket{0,1}\bra{0,1}\\
-ie^{i\Delta\phi}\ovybin \ket{0,1}\bra{1,0}\\
+ie^{-i\Delta\phi}\ovinyb \ket{1,0}\bra{0,1})
\end{split}
\tag{S8}
\end{equation}
where we have used the relations $\ovinin=\ovybyb=1$ and $\ovybin=\ovinyb^{*}=\sum_{\omega}^{} \xi^{*}\uin[,\omega]\xi\uyb[,\omega]$.

Finally, we compute the fidelity using the target state $\ket{\Psi_\text{ent}}=\frac{1}{\sqrt{2}}(\ket{1,0}-ie^{i\Delta\phi}\ket{0,1})$
\begin{equation}
F=\bra{\Psi_\text{ent}}  \rho^{\text{Yb,In}} \ket{\Psi_\text{ent}}=\frac{1}{2}\Big[1+\text{Re}(\ovybin)\Big]
\tag{S9}
\end{equation}

\subsection{Double Excitations}

Here we will derive the parameter $c_1$ of Eq.~5 of the main text. In an experimental set-up, when a photon is detected on either detector, there is a non-zero probability that both qubits were excited but only one was detected. This probability is given by
\begin{equation}
\begin{split}
    p_\text{double}
    = & [p\uybn[1,]p\uinn[1,]p\uybn[2,](1-p\uinn[2,])] \\ 
    + & [p\uybn[1,]p\uinn[1,]p\uinn[2,](1-p\uybn[2,])]
\end{split}
\tag{S10}
\end{equation}
where the two terms come from the probability of detecting one photon from either qubit that has decayed from its excited state. There is a phase of $\pi/2$ between these two photons as with the state in Eq. S6, and an additional phase factor determined by the total optical path length of the In system $\phi_{D_2}=ikD_2$.
Following through the same process, we arrive at an entangled state 

\begin{equation}
\begin{split}
\ket{\Psi_{1,2}}=\frac{1}{\sqrt{2+c_{1}^2}} \Bigg[
\ket{1,0}\sum_\omega \xi\uyb[,\omega] \frac{\cyb+i\dyb}{\sqrt{2}} \vac\\
-i e^{i\Delta\phi}\ket{0,1}\sum_{\omega^{'}}^{} \xi\uin[,\omega^{'}] \frac{\din+i\cin}{\sqrt{2}} \vac  ]\\
+c_{1}\ket{1,1} \Big[ \sum_\omega \xi\uyb[,\omega] \frac{\cyb+i\dyb}{\sqrt{2}} \vac\\-ie^{i(\phi_{D_2})}\sum_{\omega^{'}}^{} \xi\uin[,\omega^{'}] \frac{\din+i\cin}{\sqrt{2}} \vac \Big]\Bigg]
\end{split}
\tag{S11}
\end{equation}
\newpage
\hspace{-8pt}where we have that
\begin{equation}
\begin{split}
c_{1} 
    = & \frac{\sqrt{p_{double}}}{\sqrt{p\uybn[1,](1-p\uinn[1,])p\uybn[2,]}}\\
    = & \frac{\sqrt{p\uinn[1,](p\uybn[2,](1-p\uinn[2,])+p\uinn[2,](1-p\uybn[2,]))}}{\sqrt{(1-p\uinn[1,])p\uybn[2,]}}.
\end{split}
\tag{S12}
\end{equation}

Since the target state has no $\ket{1,1}$ component, when we calculate the fidelity, we obtain the same result as before, with the only modification being the prefactor $\frac{1}{\sqrt{2}}\rightarrow\frac{1}{\sqrt{2+c_{1}^{2}}}$.

Including the effect of $F_{dyn}$ from Cabrillo \textit{et al.}\cite{cabrillo1999ces} on photon distinguishibility, we then obtain the fidelity equation found in the main text:
\begin{equation}
F=\frac{1}{2+c_{1}^{2}}(1+F\utxt{dyn}\text{Re}(\ovybin).
\tag{S13}
\end{equation}

\vspace{3pt}

\end{document}